\documentclass{PoS}

\title{Search for Neutrinos from Populations of Optical Transients}

\ShortTitle{Neutrinos from Optical Transient Populations}

\author{
The IceCube Collaboration\footnote{For collaboration list, see PoS(ICRC2019) 1177.}\\
{\itshape \href{http://icecube.wisc.edu/collaboration/authors/icrc19_icecube}{http://icecube.wisc.edu/collaboration/authors/icrc19\_icecube}}\\
E-mail: \email{robert.stein@desy.de}
}

\abstract{

Since the detection of high-energy cosmic neutrinos at the IceCube Neutrino Observatory in 2013, there has been an on-going search to find the origins of this flux. Despite recent evidence identifying a flaring blazar as a possible neutrino source, the vast majority of the diffuse neutrino flux measured by IceCube remains unexplained. Here, the latest IceCube results testing time-dependent correlation between neutrinos and Tidal Disruption Events (TDEs) are presented, limiting the contribution of jetted and non-jetted TDEs to the diffuse astrophysical neutrino flux to be less than 1.3\% and 26\% respectively at 90\% confidence level. In addition, a dedicated search for neutrinos from the extraordinary transient AT2018cow are presented, and upper limits on the integrated neutrino emission are derived. Expected improvements from new and upcoming time domain optical surveys (such as ZTF and LSST) are also introduced.\\

\vspace{4mm}
{\bfseries Corresponding authors:}
\speaker{Robert Stein}$^{1}$\\
{$^{1}$ \itshape DESY Zeuthen, Platanenallee 6, 15738 Zeuthen, Germany}\\

}

\FullConference{36th International Cosmic Ray Conference -ICRC2019-\\
		July 24th - August 1st, 2019\\
		Madison, WI, U.S.A.}

\begin{document}
	
\section{Introduction}

The IceCube Neutrino Observatory is a cubic-kilometer array buried 1.5 km beneath glacier ice at the geographic South Pole \cite{Aartsen:2016nxy}. When neutrinos undergo charged-current  or neutral-current interactions in the ice, daughter charged leptons emit Cherenkov light that can be detected by IceCube's 5160 Digital Optical Modules (DOMs). In 2013, IceCube discovered a diffuse flux of high-energy astrophysical neutrinos \cite{Aartsen:2013jdh}, and there has since been an ongoing search to find source candidates. The consistency of this flux with an isotropic distribution suggests that it has a predominantly extragalactic origin. While untargeted analyses can be used to find clustering solely using neutrino data, targeted searches using multi-wavelength and multi-messenger data are employed to seek an excess of neutrinos correlated with a given source or source class. In general, the sensitivity of neutrino telescopes is limited by the background flux of atmospheric neutrinos, as well as atmospheric muons, which exceed the measured astrophysical neutrino flux by orders of magnitude except at very high energies above O(100TeV). 

This background can be overcome with two complementary approaches. In the neutrino-driven approach, neutrino events are selected which have a high-probability to be of astrophysical origin, based on their reconstructed topology. For these neutrinos, possible counterparts can be identified. Such an approach forms the basis of the IceCube Realtime Program \cite{Aartsen:2016lmt}, in which likely astrophysical neutrinos are identified in real-time and immediately distributed as "alerts" to astronomers via the Gamma-ray Coordination Network (GCN) framework. Because only a handful of neutrinos are identified with these filters each year, it can be hard to make statistically-significant statements about source populations using this approach. The low alert rate is compounded by the abundance of undetected neutrino counterparts that would be expected for most source populations. For a population of neutrino sources similar to Core-Collapse Supernovae (CCSNe), just 20\% of all astrophysical neutrinos would be expected to have detectable counterparts \cite{Kankare:2019bzi}, assuming that they were followed up by the most sensitive current instruments.

In the alternative source-driven approach, specific source hypotheses are tested. These searches typically exploit the large expected number of lower-energy astrophysical neutrinos, enabling analysis with significantly higher statistics at the cost of greater atmospheric background \cite{Aartsen:2016oji}. Requiring spatial coincidence with a potential source does, however, significantly reduce the background for a search. As astrophysical neutrino sources are expected to have significantly harder spectra than the background atmospheric neutrino flux, the incorporation of energy-dependent weighting provides additional separating power. Another effective method is to additionally require temporal coincidence, either with the lifetime of a transient, or during  pre-defined "interesting periods" for variable objects.  Multiple sources can be combined in a stacking analysis, which are designed to detect the sum of signals from many weak individual sources. In all cases, these methods rely on multi-messenger and multi-wavelength observations to pre-identify sources to be analysed.

The most successful example of the neutrino-driven approach came with the detection of the high-energy neutrino IC170922A, which launched a comprehensive multi-messenger follow-up campaign \cite{IceCube:2018dnn}. The Fermi collaboration reported that the neutrino was coincident with a flaring gamma-ray blazar. A chance coincidence of this kind was disfavoured at the level of 3$\sigma$. An archival analysis unexpectedly revealed an additional 3$\sigma$ excess of neutrinos at the position of this same blazar in 2014, during a period for which there was no significant gamma-ray flaring activity \cite{IceCube:2018cha}. At the same time, previous IceCube analyses have limited the cumulative distribution of Fermi 2LAC blazars to the astrophysical neutrino flux to be less than 30\% \cite{Aartsen:2016lir}. The origin of the vast majority of the diffuse neutrino flux thus remains, as yet, undiscovered. Dedicated searches targeting likely sources, including Gamma-Ray Bursts (GRBs), CCSNe, Starburst galaxies, and galactic emission, have so far failed to reveal any significant excess above background expectations \cite{Stasik2018Search}. This motivates the continued analysis of new, untested source classes in an attempt to identify the origin of astrophysical neutrinos. Within this context, a new analysis was undertaken to search for neutrinos from Tidal Disruption Events (TDEs).

\section{Tidal Disruption Events}

A TDE occurs when a star approaches a supermassive black hole (SMBH) on a parabolic orbit \cite{Komossa:2015qya}. As gravitational acceleration follows a $\frac{1}{r^{2}}$ dependence, the near side of the star will be accelerated more strongly than the far side. The star thus experiences a net tidal force. As the star moves closer to the SMBH, the tidal force increases, until it exceeds the self-gravity that holds the star together. At this point, the star is said to be tidally-disrupted, and roughly half of the stellar debris is accreted. In some cases, a relativistic jet can be formed during the accretion process, analogously to a blazar jet. There has been recent theoretical interest in TDEs as potential Ultra-High Energy Cosmic Ray (UHECR) sources, as well as candidate neutrino sources, see e.g \cite{Lunardini:2016xwi, Biehl:2017hnb}.

TDEs are a fundamentally rare phenomenon, with rates several orders of magnitude below CCSN rates \cite{vanVelzen:2017qum}. However, historically poor detection efficiencies have further exacerbated this, leaving only a handful of reliably-identified TDEs. To date, there have been only 3 on-axis jetted TDEs, and a few dozen candidate non-jetted TDEs \cite{Komossa:2015qya, Auchettl:2016qfa}. Among these, the majority do not have an unambiguous TDE classification. 

TDEs themselves are, by their nature, nuclear transients. They can often be confused with flares of Active Galactic Nuclei (AGN), as well as nuclear CCSNe. Due to the greater abundance of these background populations, it can be hard to remove all contamination. Ultimately, multiple eras of spectroscopy and photometry are required for a compelling classification. At the time of catalogue compilation in October 2017 \cite{Auchettl:2016qfa}, out of approximately 60 candidate TDEs in the literature overlapping the IceCube data-taking period, only 13 were judged to be unambiguously classified.

\section{Analysis}

Owing to the diverse nature of TDEs observed to date, spanning a range of luminosities and multi-wavelength evolution, a traditional stacking analysis using a fixed weighting scheme would not be well-tailored to searching for neutrino emission. While such searches are optimal for cases in which the expected intrinsic neutrino luminosity of each source is known, it quickly becomes less sensitive to deviations from the assumed hypothesis. Observational diversity and uncertainty, coupled to uncertainty in theoretical model predictions, mean that robust predictions for the individual neutrino luminosities of each source cannot be made. Instead, a more agnostic stacking method was employed for the analysis that did not make any assumptions on the relative strength of each tested source, and was thus robust against both catalogue contamination and deviations from a standard-candle neutrino emission scheme \cite{Stasik2018Search}. This unbinned likelihood analysis method was applied to 9.5 years of muon neutrino data, extending from April 2008 to October 2017 \cite{IceCube:2018cha}. Despite the agnostic search method, in order to meaningfully interpret the results and extrapolate to constrain emission from the population as a whole, a pure TDE sample is still required. 

Consequently, the non-jetted sample was separated based on robustness of classification. A "golden sample" of unambiguous TDEs was created, and a separate "silver sample" of likely TDEs was also created. A third category of "obscured TDEs" was also created, for Infra-Red (IR) flare TDE candidates in dusty galaxies \cite{Wang:2018mxl}. As these obscured TDEs were detected only via IR observation, for which we expect significant reprocessing, there would likely be a variable time lag between the disruption itself and the corresponding IR flare. As we expect that neutrinos should generally arrive after the disruption, these obscured TDEs were treated with larger, more conservative search windows to account for this additional uncertainty. These three categories do not form physically distinct classes, but rather differing levels of observational data and likely contamination.

For each source, a tailored search window was defined, with the aim of covering the period from 30 days before peak to 100 days after. This specific choice was motivated by the range of theoretical predictions for TDE neutrino emission, all of which broadly agree that emission should coincide with the period of peak electromagnetic brightness following the disruption. Due to the sparsity of available observational data, this search window was conservatively extended for sources without a resolved lightcurve peak, extending to the last available upper limit. In cases where no upper limit was available, the window was instead set to be one year before observed peak up to 100 days afterwards. With this method, there is reasonable certainty that the lightcurve peak fell within the search window for each TDE. For the obscured TDEs, accounting for time lags that were expected to be roughly of order 100 days, all windows were fixed to extend from one year before IR lightcurve peak, to 100 days after peak.

\begin{table}[]
    \centering
     \begin{tabular}{||c c c c||} 
     \hline
     Catalogue & Source Class & Size & Description \\ [0.5ex] 
     \hline\hline
     Jetted & Jetted TDEs &  3 & \textit{Probable TDEs with on-axis jets}\\ 
     \hline
     Golden & Non-Jetted TDEs & 13 & \textit{Probable TDEs with convincing classification}\\
     \hline
     Silver & Non-Jetted TDEs & 24 & \textit{Candidate TDEs with ambiguous classification}\\
     \hline
     Obscured & Non-Jetted TDEs & 13 & \textit{Candidate TDEs in dusty galaxies}\\[1ex] 
     \hline
     \end{tabular}
    \caption{Summary of the four TDE catalogues. For each, an independent stacking analysis was performed.}
    \label{tab:my_label}
\end{table}{}

\section{Results}

For each of the four catalogues, an independent stacking analysis was performed. In all cases, the results were consistent with expectations from background, and thus upper limits are accordingly derived at 90\% confidence level. Separate upper limits were derived for the two distinct source populations, namely on-axis jetted TDEs and non-jetted TDEs. For the calculation of these limits on source population, the "Golden TDEs" were assumed to be representative of non-jetted TDEs as a whole. 

\begin{figure}[!ht]
	\centering \includegraphics[width=\textwidth]{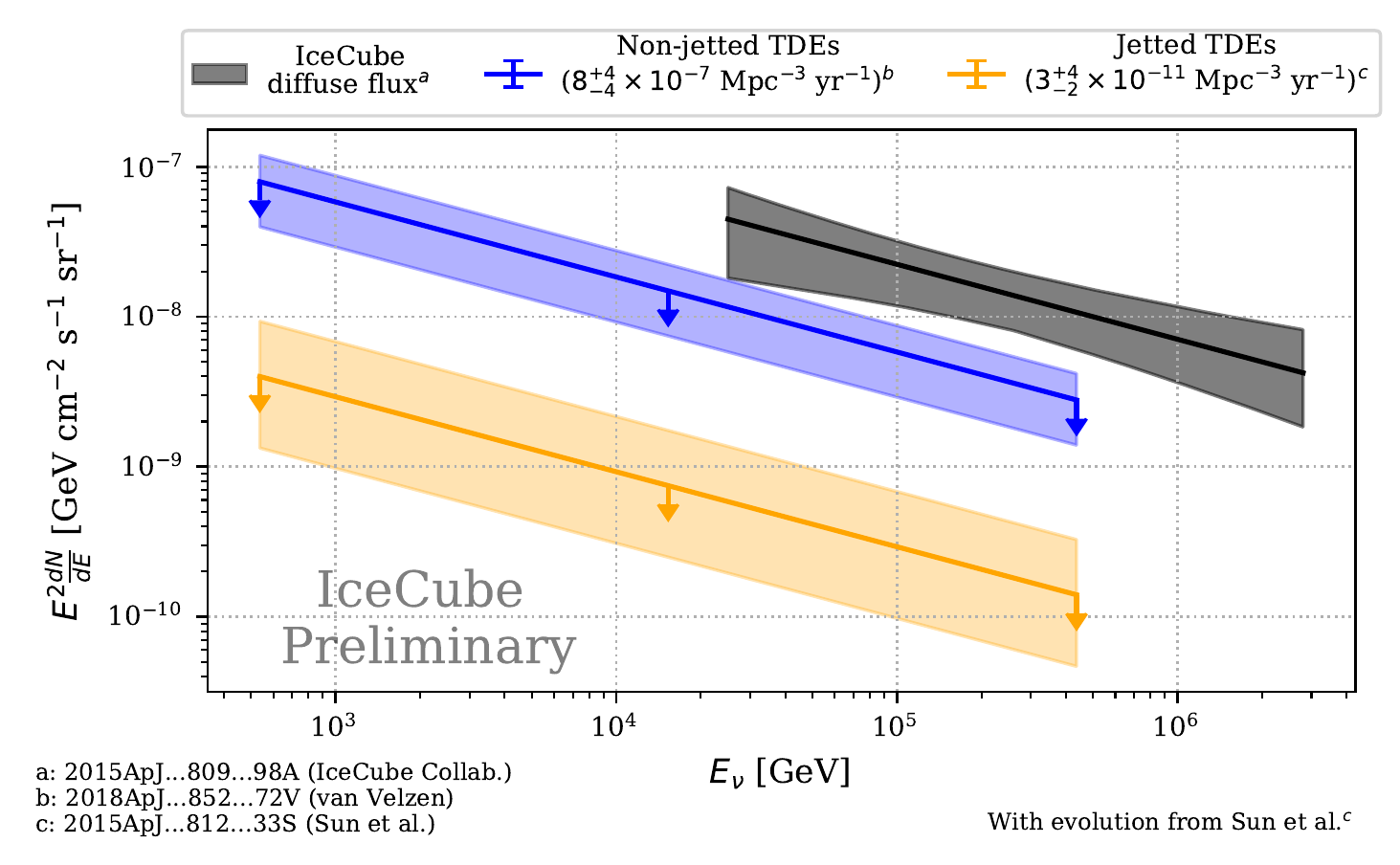}
	\caption{90\% confidence level upper limits on the contribution of jetted and non-jetted TDEs to the diffuse neutrino flux \cite{Aartsen:2015knd}, assuming standard candle behaviour. The shaded bands represent uncertainty in local rate estimates of TDEs from \cite{vanVelzen:2017qum, Sun:2015bda}}
	\label{fig:DiffuseFlux}
\end{figure}

By assuming that these TDEs behave as standard candles, source class limits on neutrino emission can be derived.  The results are shown in Figure \ref{fig:DiffuseFlux}. Assuming the central value of rate estimates from \cite{vanVelzen:2017qum} and \cite{Sun:2015bda}, and an $E^{-2.5}$ astrophysical neutrino flux, we find that non-jetted and jetted TDEs contribute less than 26\% and 1.3\% respectively to the astrophysical neutrino flux. As the contribution from a population is directly proportional to the local population rate, the shaded bands indicate the uncertainty in our limits arising from rate estimates. For TDEs, these rates are the dominant source of uncertainty in neutrino flux constraints. It will require systematic evaluation of observed TDE rates to enable more precise limits on neutrino emission. Any refined rate estimate can be immediately used to directly recalculate  limits, without requiring any additional IceCube analysis.

\begin{figure}[!ht]
	\centering \includegraphics[width=\textwidth]{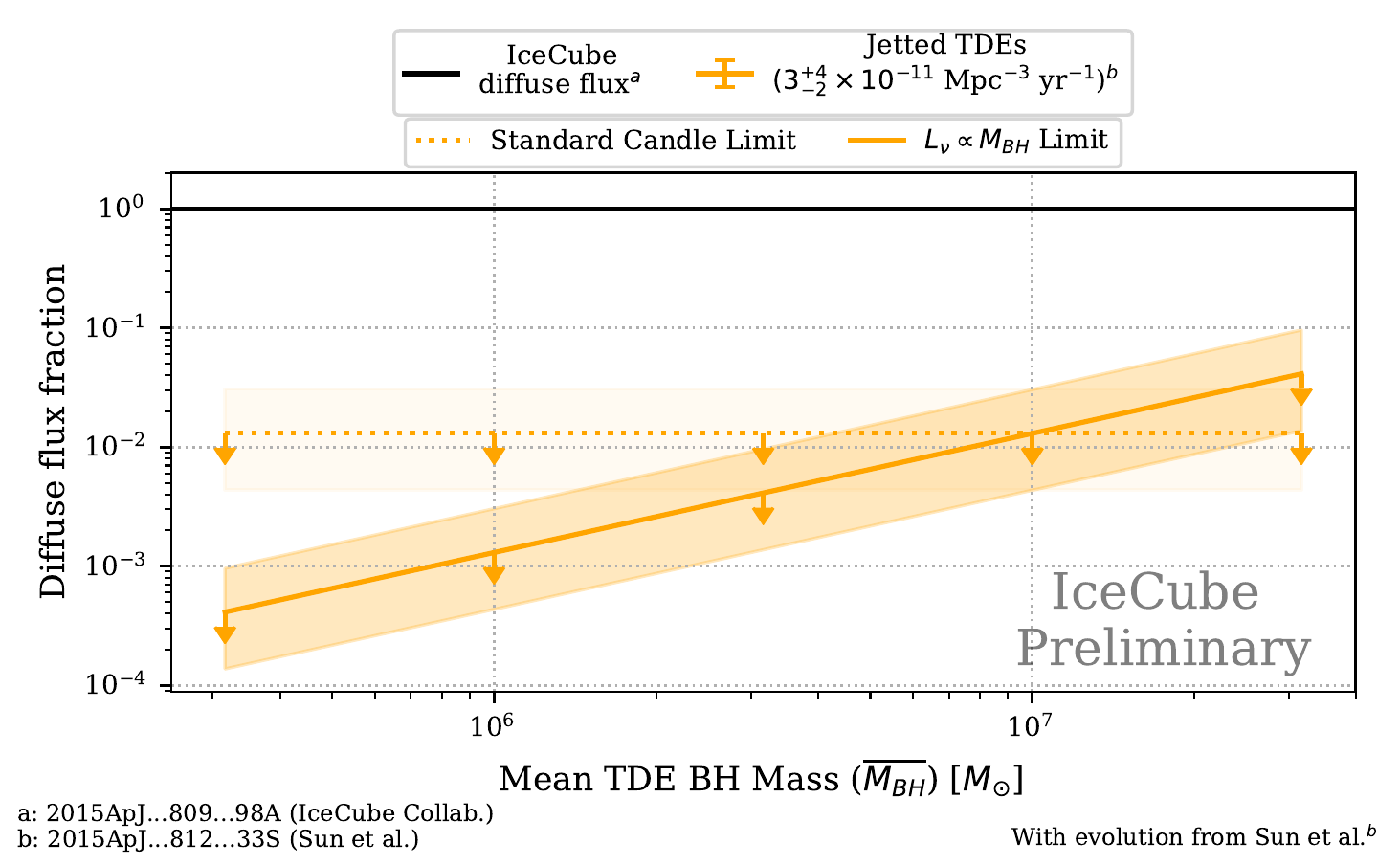}
	\caption{90\% confidence level upper limits on the contribution of jetted  TDEs to the diffuse neutrino flux \cite{Aartsen:2015knd} as a function of mean TDE SMBH mass, assuming either standard candle behaviour or $L_{nu} \propto M_{BH}$. The contribution to the neutrino flux is directly proportional to the assumed mean TDE SMBH mass for the $L_{nu} \propto M_{BH}$, but is completely independent for the standard candle case. The shaded bands represent uncertainty in local rate estimates of TDEs from \cite{Sun:2015bda}}
	\label{fig:DiffuseFluxMass}
\end{figure}

An alternative hypothesis was tested for Jetted TDEs, in which the neutrino luminosity was assumed to be proportional to the SMBH mass. This assumption was motivated by the Eddington Limit, which limits the accretion and is proportional to black hole mass. Observational evidence further suggests that TDE bolometric luminosities do tend to broadly follow such a relation \cite{wevers}. In this case, the limits are directly proportional to the mean SMBH mass for the TDE population, as illustrated in Figure \ref{fig:DiffuseFluxMass}. This mean mass was assumed to be $10^{6.5} M_{\odot}$, a value consistent with observations of TDE hosts \cite{wevers}. Under these assumptions, the contribution of jetted TDEs to the diffuse neutrino is then limited to less than 0.4\% of the total. 

\section{AT2018cow}

The discovery of extraordinary transient AT2018cow was a further demonstration of the central importance of optical telescopes for identifying transients.  This fast, bright, blue transient prompted a comprehensive multi-messenger follow-up campaign. The observations were consistent with a nearby example of a recently-identified population of Fast Blue Optical Transients (FBOTs), see e.g \cite{Margutti:2018rri}. 

Shortly after the time of discovery, AT2018cow was thought to be a Broad-Lined type Ic (Ic-BL) supernova, and thus a member of the rare CCSN subclass associated with long GRBs and choked-jets. As many models predict that such SNe may be neutrino sources, an IceCube Fast Response Analysis was performed on AT2018cow shortly after discovery \cite{icrc_19_fra}. Within the context of a candidate choked-jet supernova, the IceCube search spanned the 3-day period from the last non-detection to the first detection, aiming to isolate the supernova explosion time at which the neutrino emission would be expected. Ultimately, an excess of neutrinos was found in this time period, with a significance of 1.8 $\sigma$, and the results of the search were published through the Astronomers Telegram network \cite{2018ATel11785....1B}. The excess itself consisted of two well-reconstructed neutrinos, which were considered significant owing to the small expected background for such a short search window.

\begin{figure}[!ht]
	\centering \includegraphics[width=.9\textwidth]{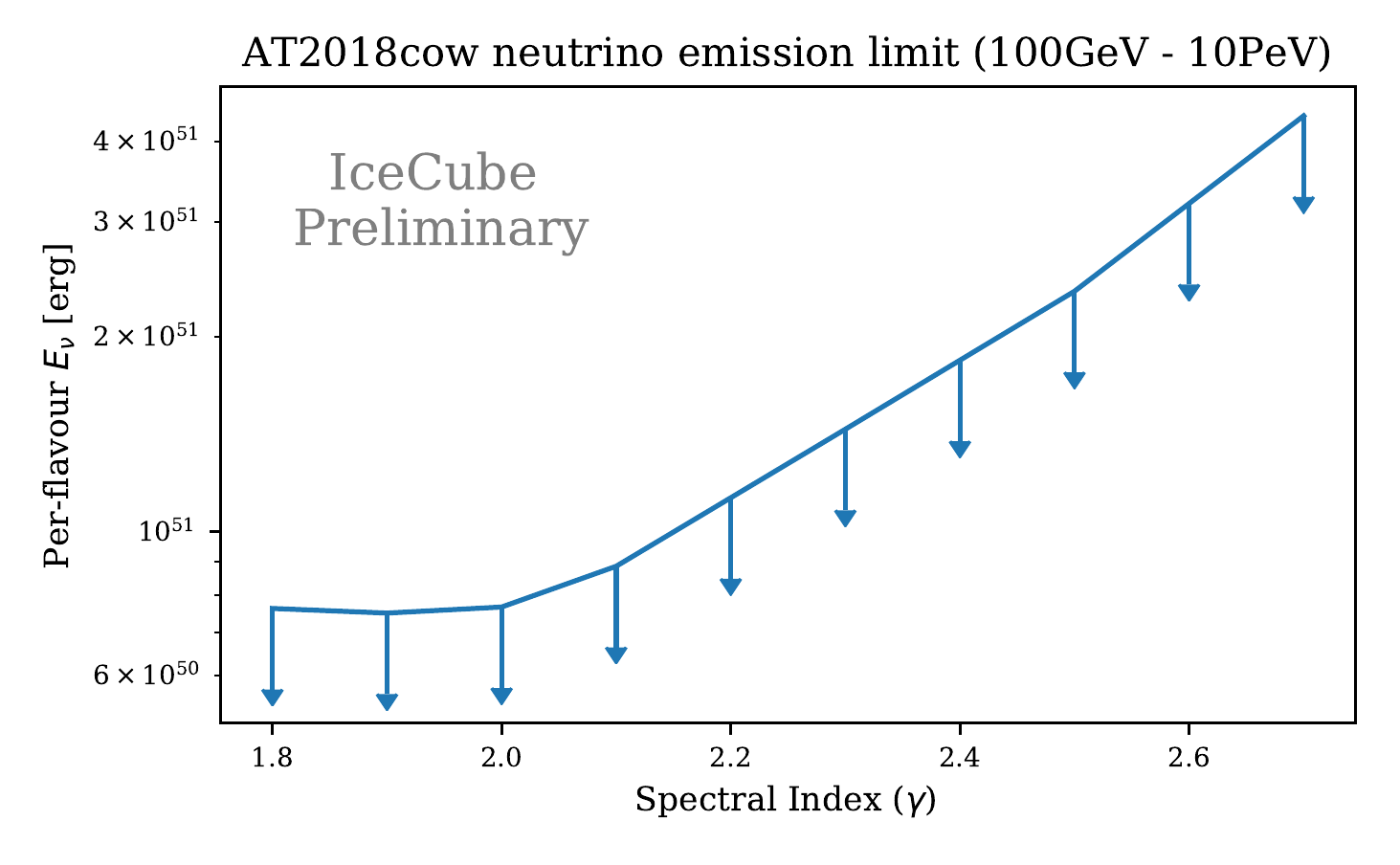}
	\caption{90\% confidence level upper limit on integrated neutrino emission from AT2018cow as a function of spectral index, assuming a 130 day window from MJD 58256.9 to MJD 58386.9}
	\label{fig:At2018cow}
\end{figure}

Later multi-wavelength observations of AT2018cow were not consistent with a traditional Ic-BL SN, and the transient has since variously been interpreted as a TDE, an extreme SN or a Magnetar \cite{Perley:2018oky}. In light of these developments, AT2018cow was re-analysed by IceCube in the context of a potential TDE classification. A dedicated search for neutrino clustering on variable timescales was performed \cite{IceCube:2018cha}. For TDEs, these timescales were restricted to a maximum of 130 days, extending from 30 days before peak to 100 days afterwards. A small neutrino excess was found, with the best-fit cluster including the same neutrino events that were found in the original IceCube analysis. However, when accounting for the expected fluctuations arising from background over the much longer 130 day search window, the significance of the excess was just 0.5 $\sigma$. The result is thus entirely consistent with expectations from atmospheric background, and is also compatible with the result of the original IceCube analysis. As such, no evidence of neutrino emission is claimed and a 90\% confidence upper limit is derived accordingly (illustrated in Figure \ref{fig:At2018cow}). As before, uncertainty in both classification and rate estimates hinder attempts to constrain neutrino emission from FBOTs.

\section{Summary and Outlook}

A search was conducted looking for the first time for evidence of neutrino emission from both jetted and non-jetted TDEs. No such evidence was found, indicating that jetted and non-jetted TDEs contribute less than 1.3\% and 26\% respectively to the astrophysical neutrino flux, assuming the rates in \cite{vanVelzen:2017qum, Sun:2015bda}. As an emerging transient class which is still not well understood, there remains significant uncertainty in the rates of TDEs, and this directly translates into uncertainty in the cumulative population flux. 

Fortunately, the emergence of new facilities such as ZTF \cite{2019PASP..131a8002B}, as well as future surveys such as LSST, should lead to gains for all optical transient neutrino analyses. By discovering greater numbers of transients, the sensitivity of searches will grow. Larger samples should also improve rate estimation, in particular for TDEs where no large systematic sample has been collated before. Higher cadence observations can greatly reduce background by constraining search windows, for example the estimated CCSN explosion time, with greater precision. Consequently, source-driven analysis will continue to grow more powerful.

\bibliographystyle{ICRC}
\bibliography{my-bib-database}

%

\end{document}